\documentclass{article}

\usepackage{arxiv}

\usepackage{amsmath}
\usepackage[utf8]{inputenc} 
\usepackage[T1]{fontenc}    
\usepackage{hyperref}       
\usepackage{url}            
\usepackage{booktabs}       
\usepackage{amsfonts}       
\usepackage{nicefrac}       
\usepackage{microtype}      
\usepackage{cleveref}       
\usepackage{lipsum}         
\usepackage{graphicx}
\usepackage{doi}
\usepackage{physics}
\usepackage{xcolor}
\usepackage[ruled, linesnumbered]{algorithm2e}
\usepackage{tcolorbox}
\usepackage[frozencache,cachedir=.]{minted}
\tcbuselibrary{minted,breakable,xparse,skins}
\usepackage{caption}
\usepackage{subcaption}
\usepackage{dcolumn}
\usepackage{bm}
\DeclareCaptionLabelFormat{codecap}{Code #2} 
\newenvironment{code}{\captionsetup{type=listing,labelformat=codecap}}{} 
\definecolor{bg}{gray}{0.95}
\DeclareTCBListing{mintedbox}{O{}m!O{}}{ 
  breakable=true,
  listing engine=minted,
  listing only,
  minted language=#2,
  minted style=default,
  minted options={%
    linenos,
    gobble=0,
    breaklines=true,
    breakafter=,,
    fontsize=\small,
    numbersep=8pt,
    #1},
  boxsep=0pt,
  left skip=0pt,
  right skip=0pt,
  left=25pt,
  right=0pt,
  top=3pt,
  bottom=3pt,
  arc=5pt,
  leftrule=0pt,
  rightrule=0pt,
  bottomrule=2pt,
  toprule=2pt,
  colback=bg,
  colframe=orange!70,
  enhanced,
  overlay={%
    \begin{tcbclipinterior}
    \fill[orange!20!white] (frame.south west) rectangle ([xshift=20pt]frame.north west);
    \end{tcbclipinterior}},
  #3}

\title{Image reconstruction from an elastically distorted scan}

\date{}

\newif\ifuniqueAffiliation

\ifuniqueAffiliation 
\author{ \href{https://orcid.org/0000-0000-0000-0000}{\includegraphics[scale=0.06]{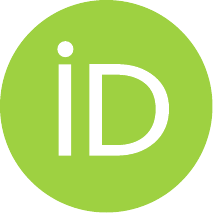}\hspace{1mm}David S.~Hippocampus}\thanks{Use footnote for providing further
		information about author (webpage, alternative
		address)---\emph{not} for acknowledging funding agencies.} \\
	Department of Computer Science\\
	Cranberry-Lemon University\\
	Pittsburgh, PA 15213 \\
	\texttt{hippo@cs.cranberry-lemon.edu} \\
	\And
	\href{https://orcid.org/0000-0000-0000-0000}{\includegraphics[scale=0.06]{orcid.pdf}\hspace{1mm}Elias D.~Striatum} \\
	Department of Electrical Engineering\\
	Mount-Sheikh University\\
	Santa Narimana, Levand \\
	\texttt{stariate@ee.mount-sheikh.edu} \\
}
\else
\usepackage{authblk}

\newbox{\orcid}\sbox{\orcid}{\includegraphics[scale=0.06]{orcid.pdf}} 
\author[1]{%
	\href{https://orcid.org/0000-0002-7915-2597}{\usebox{\orcid}\hspace{1mm}Adrian Lopez}%
}
\author[2,3]{%
	\href{https://orcid.org/0000-0002-5114-0519}{\usebox{\orcid}\hspace{1mm}L. Mahadevan\thanks{\texttt{lmahadev@g.harvard.edu}}}%
}
\affil[1]{Graduate School of Arts and Sciences, Harvard University, Cambridge, Massachusetts 02138, USA}
\affil[2]{School of Engineering and Applied Sciences, Harvard University, Cambridge, Massachusetts 02138, USA}
\affil[3]{Departments of Physics, and of Organismic and Evolutionary Biology, Harvard University, Cambridge, Massachusetts 02138, USA}
\fi

\renewcommand{\headeright}{}
\renewcommand{\undertitle}{}
\renewcommand{\shorttitle}{}

\hypersetup{
pdftitle={Image reconstruction from an elastically distorted scan},
pdfauthor={Adrian Lopez, L. Mahadevan},
}

\begin{document}

\maketitle

\begin{abstract}
	We consider the problem of inverting the artifacts associated with scanning a page from an open book, i.e. "xeroxing." The process typically leads to a non-uniform combination of distortion, blurring and darkening owing to the fact that the page is bound to a stiff spine that causes the sheet of paper to be bent inhomogeneously. Complementing purely data-driven approaches, we use knowledge about the geometry and elasticity of the curved sheet to pose and solve a minimal physically consistent inverse problem to reconstruct the image. Our results rely on 3 dimensionless parameters, all of which can be measured for a scanner, and show that we can improve on the data-driven approaches. More broadly, our results might serve as a "textbook" example and a tutorial of how knowledge of generative mechanisms can speed up the solution of inverse problems.
\end{abstract}

\section{Motivation}
\begin{figure*}[t!]
    \centering
    \begin{subfigure}[t]{0.9\textwidth}
        \centering
        \includegraphics[height=1.2in]{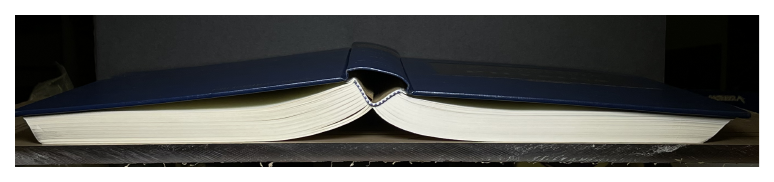}
        \caption{}
        \label{fig:book_scanner}
    \end{subfigure}
    \begin{subfigure}[t]{0.46\textwidth}
        \centering
        \includegraphics[height=2.0in]{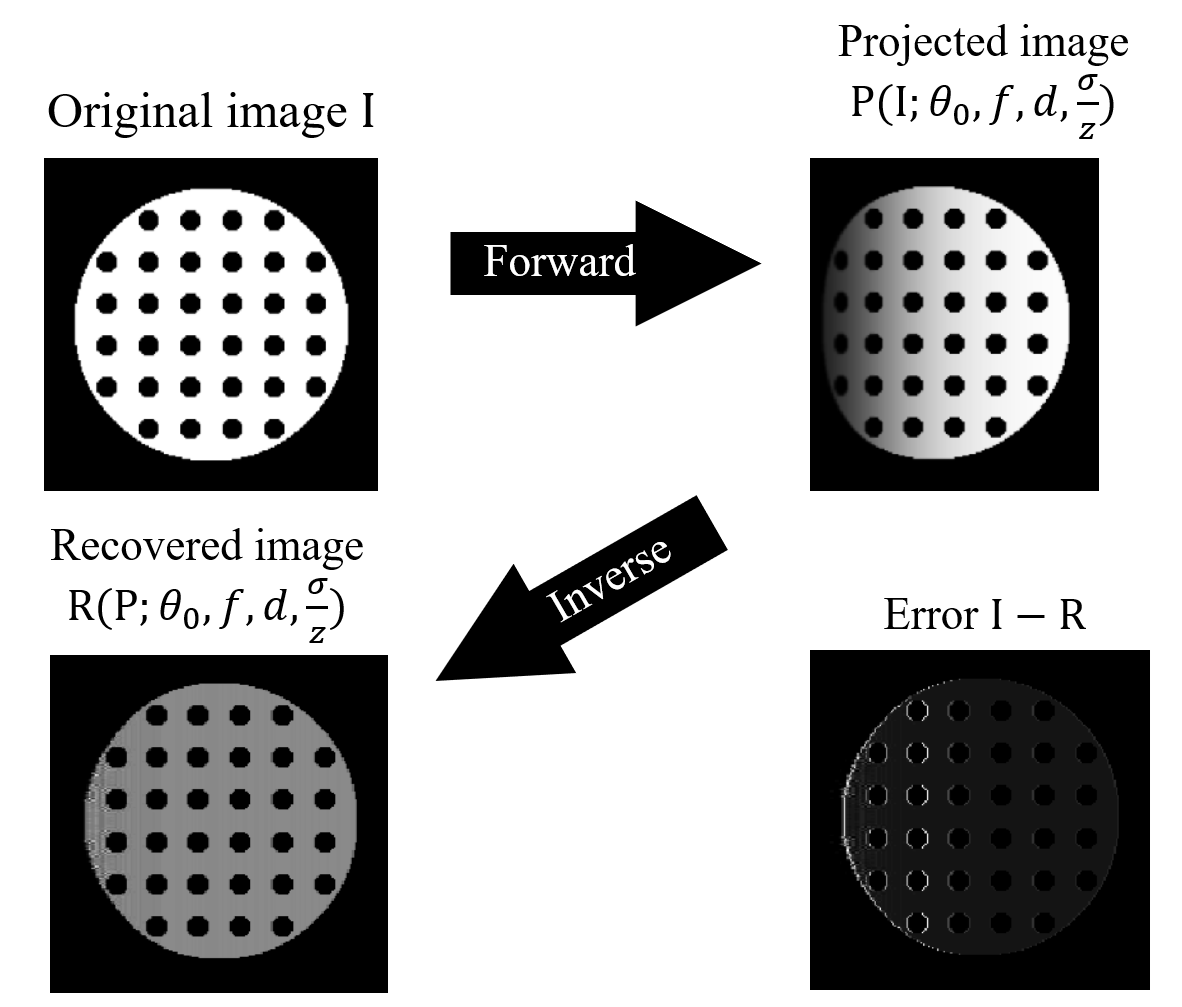}
        \caption{}
        \label{fig:book_scan}
    \end{subfigure}
    \begin{subfigure}[t]{0.46\textwidth}
        \centering
        \includegraphics[height=1.8in]{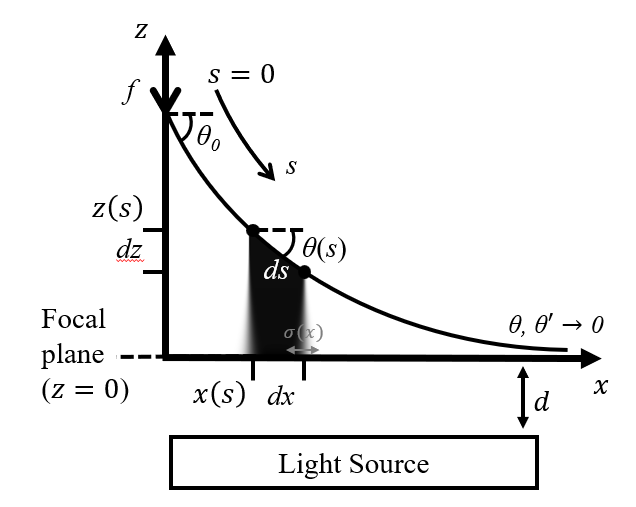}
        \caption{}
        \label{fig:book_membrane}
    \end{subfigure}
    \caption{a) A book in position to be scanned, with its pages curved near its spine. b) The forward and inverse problems applied to a simple image, showing the simulation of the three artifacts we consider as well as their correction. Note that the recovered image and error are shown with exaggerated contrast. c) The shape of the elastic membrane $(x(s), z(s))$ according to the physical model (equation \ref{x_s}-\ref{z_s}). The $y$-axis points out of the plane, and the membrane has no curvature in this direction. Following the projection of a length element elucidates the origin of the three artifacts mentioned: distortion due to the difference between the length elements $ds$ and $dx$ (section \ref{forward_dist}), blurring due to the distance $z(s)$ from the focal plane assuming linear optics characterized by $\frac{\sigma}{z}$ (section \ref{forward_blur}), and darkening due to the distance $d+z(s)$ from the light source (section \ref{forward_dark}).}
\end{figure*}

Since the beginning of advanced scanning technologies in the 1980s with the advent of ``Xerox'' machines, how many books or articles in journals might have been scanned? Rough estimates vary by orders of magnitude, but even the most conservative ones suggest that more than a trillion book pages have been scanned. The process is straightforward - open the book to the page under question, place it on the flat screen of a photocopier, and hit a button. Because the page is bound to a stiff spine, it can never be flattened completely.   Therefore, some parts of the page will lie a certain distance from the scanner's platen, and thus also the scanner's light source and focal plane. As a result, even though the process of creating an image has varied and improved over time, starting with static photographs in the early days to moving wands of light in modern scanners, the resulting image is not a perfect copy of the original. The projection suffers from a combination of inhomogeneous distortion proportional to the slope of the page, blurring due to non-uniform distance of the page from the focal plane, and darkening due to the fact that the distance from the light source to page also varies (fig. \ref{fig:book_scanner}, \ref{fig:book_scan}). These three artifacts (differential distortion, blurring, and darkening) are generally undesirable, and one may desire to invert such artifacts or ``restore'' the scan to recover the original image. Such restoration techniques result in better OCR outcomes for images of text \cite{nagy_twenty_2000} as well as better viewing for non-textual documents.

Prior work has approached this problem in various ways, including text-line detection methods \cite{ogorman_document_1993, zhang_correcting_2003, ulges_document_2005, stamatopoulos_goal-oriented_2011, kim_document_2015}, word matching \cite{kluzner_page_2011}, 3D shape reconstruction methods \cite{wu_captured_2015}, and illumination detection \cite{courteille_shape_2007, brown_restoring_2007, zhang_improved_2008}, which ultimately allow for characterization of the page shape and thus restoration. These models are essentially data-driven, employing machine learning techniques to correct scans based on high-level priors. Here we take a complementary view based on our knowledge of the generative physical process associated with how a thin sheet deforms under applied forces, known since the work of Euler~\cite{love_treatise_2013}, along with a minimal model for the scanning process.

\section{The shape of an open book page} \label{Physical Model}
We model a book page as a thin elastic sheet of thickness $h$ that is much smaller than the other dimensions of the page; typically the thickness of a page is $O(50)\; \mu$m, while the other two dimensions are $O(10)\;$cm. Therefore its deformations may be well modeled as that associated with a developable surface \cite{audoly_elasticity_2010}. Furthermore, since the sheet is bound to a spine along one edge while its other edges are free, it must assume a cylindrical form with varying (one-dimensional) curvature only in the direction transverse to the spine. The shape of the page can thus be parameterized by the angle it makes with the horizontal, $\theta(s)$, as a function of the arc length distance $s$ from its binding (fig. \ref{fig:book_membrane}). 

The spatial variation in $\theta(s)$ is determined by the force balance on the sheet when the book spine is pressed downwards towards the scanning platen. Assuming that the vertical applied force on a page at the spine $f$ is much larger than the weight of the book, we can ignore the effects of gravity. Then force balance on the elastic sheet (or equivalently, minimization of the sum of mechanical potential energy and elastic bending energy) yields a second-order differential equation for $\theta(s)$, known to and analyzed by Euler~\cite{love_treatise_2013} nearly two centuries ago,

\begin{equation}
\label{force_balance}
    B \theta_{ss} - f \sin\theta = 0,
\end{equation}
with $B$ the bending modulus of the page, and $(\cdot)_s = d(\cdot)/ds$.  Rescaling the arc length by writing $\hat{s} = s \sqrt{f/B}$ in terms of the intrinsic length scale $l \equiv \sqrt{B/f}$ that arises from a balance between elasticity and the applied force, we can express the equilibrium equation (2.1) in dimensionless form as
\begin{equation}
\label{force_balance_nat}
    \theta_{\hat{s}\hat{s}} - \sin \theta = 0.
\end{equation}

We then apply the boundary condition $\theta(\infty)=\theta_{\hat{s}}(\infty)=0$, which physically corresponds to the page asymptotically approaching flatness (along both principal directions) far from the spine. With this condition, an integration and rearrangement of \ref{force_balance_nat} yields

\begin{equation}
        \theta_{\hat{s}} = -2 \sin{\left( \frac{\theta}{2} \right)},
\end{equation}
which is first-order and linear in $\theta_{\hat{s}}$. A second integration then gives
\begin{equation}
\label{theta_s}
    \tan{\left(\frac{\theta}{4}\right)} = \tan{\left(\frac{\theta_0}{4}\right)} e^{-\hat{s}}, 
\end{equation}
where $\theta_0 \equiv \theta(s=0)$ is the angle of the page at the spine.

This equation for $\theta(\hat{s})$ characterizes the book shape that we will assume in our subsequent  recoveries and simulated projections of scans. Explicitly, we can now find the $(x,z)$ coordinates that parameterize the membrane via 
\begin{equation}
    \label{x_s}
    x(\hat{s}) = l \int_0^{\hat{s}} d\hat{s}' \cos{\left(\theta(\hat{s}')\right)}
\end{equation}

\begin{equation}
    \label{z_s}
    z(\hat{s}) = l \int_0^{\hat{s}} d\hat{s}' \sin{\left(\theta(\hat{s}')\right)} .    
\end{equation}

\begin{figure*}
    \centering
    \includegraphics[height=2.2in]{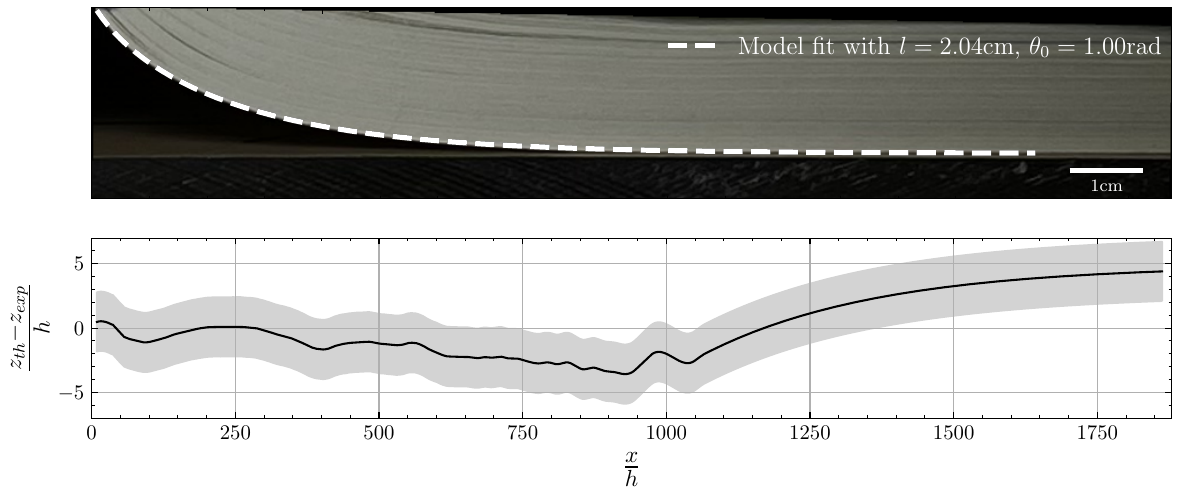}
    \caption{Fitting the physical model to an experimental parametrization of a book page's shape. The upper plot shows the result of the theoretical model with the best fit parameters, overlaid onto the relevant section of the book. The bottom plot shows the residuals between the best fit model and the experimentally extracted page positions; the gray region lies within a distance of 1 pixel from the residuals and represents the uncertainty of our experimental page shape detection. Both $x$ and $z$ are shown in units of the page thickness $h=0.067$mm; it can be seen that the model accurately capures the page shape to within a few pages. See figure \ref{fig:shape_verification_2} for a situation where this model results in higher errors due to its simplifications.}
    \label{fig:shape_verification}
\end{figure*}

Figure \ref{fig:shape_verification} shows a typical result of fitting this physical model to the page of a book placed face-down. We see a good agreement with the most systematic error occurring in regions of the page making contact with the platen, a feature which our model neglects.

\section{Projection (Forward Problem)} \label{forward}
Here we present a method for solving the ``forward problem'' of projection, that is, simulating the projection of an image printed on a page with known shape onto a flat photocopier screen below (fig. \ref{fig:forward_monet_a}, \ref{fig:forward_monet_b}). In our formulation of the forward and inverse problems, we allow for a general page shape, but in our example applications we assume the aforementioned elastic profile.

We will consider the image as a function of two variables $I(x,y)\equiv I(\vec{x})$. In discrete terms one can think of $I(x,y)$ as the value of the pixel at location $(x,y)$. We consider $I$ to be a scalar, corresponding to a monochromatic image, but note that our treatment generalizes simply to vector functions (color images); one only needs to repeatedly apply the scalar treatment for each component of the image function.

\subsection{Distortion} \label{forward_dist}
When the image on the curved page is vertically projected, the image will be locally distorted in the $x$-direction, that is, horizontally ``squeezed'' by different amounts at different locations. The distortion factor of a given segment of the image is the ratio of the arc length of the segment along the membrane to the length of its projection onto the $x$-axis. This local factor is thus given by the cosine of the angle $\theta$ between the curved membrane and flat screen, and so can be simulated by remapping the coordinates on which $I$ is defined:

\begin{align}
\label{distort}
    \textrm{Dist}\left(I\right)(x,y) &= I(\hat{s},y), \\
    \frac{dx}{d\hat{s}} &= l \cos \theta(\hat{s})
\end{align}
where $\textrm{Dist}(I)$ is the distorted version of $I$. Although there is no curvature in the  $y$-direction, there can exist distortion along this dimension due to the page's varying distance from the camera. Since this effect can be eliminated by choosing a particular scan direction, we assume in the remainder of our model that no distortion takes place along the $y$-direction.

\subsection{Blurring} \label{forward_blur}
The projected image will also be differentially blurred, with the extent of the blur of a given region depending on its distance from the focal plane. We formulate this by assuming linear optics, in which each pixel undergoes a Gaussian blur with a standard deviation proportional to its distance from the focal plane, $\sigma(x,y) \propto z(x)$. Note that we have also assumed the focal plane to be identical to the plane of the platen at $z=0$. This blur is then implemented by considering the contribution to the final image from each blurred pixel (Gaussian point-spread functions), then summing all such point-spread functions with weights according to the pixel's original value. Defining $B(\vec{x})=\textrm{Blur}(I(\vec{x}))$ as the blurred version of $I$, such a linear differential blur is given by

\begin{align}
    B(\vec{x}) &= \textrm{Blur}(I(\vec{x})) \\
    &= \textrm{Blur}\left( \int d^2 \vec{x}' \; \; I(\vec{x}') \delta^2(\vec{x}' - \vec{x})\right) \\
    &= \int d^2 \vec{x}' \; \; I(\vec{x}') \textrm{Blur}\left( \delta^2(\vec{x}' - \vec{x})\right) \\
    &= \int d^2 \vec{x}' \; \; I(\vec{x}') \frac{1}{\sigma(z(\vec{x}')) \sqrt{2\pi}} \exp\left( \frac{-(\vec{x}-\vec{x}')^2}{2 \sigma(z(\vec{x}'))^2} \right) .
\end{align}

For future convenience we note that the blur function can be expressed in terms of a linear kernel,
\begin{align}
\label{blur_kernel}
    \textrm{Blur}(I(\vec{x})) &= \int d^2\vec{x}' \; K(\vec{x},\vec{x}') I(\vec{x}') \\
    \textrm{where } K(\vec{x},\vec{x}') &= \frac{1}{\sigma(z(\vec{x}')) \sqrt{2\pi}} \exp\left( \frac{-(\vec{x}-\vec{x}')^2}{2 \sigma(z(\vec{x}'))^2} \right).
\end{align}

\subsection{Darkening} \label{forward_dark}
Due to the distance of the membrane from the light source of the scanner, the projected image will also be differentially darkened by a factor proportional to a negative power of the membrane's distance from the light source. Thus the darkened version of the image is given by 

\begin{equation}
\label{darken}
    \textrm{Dark}(I(\vec{x})) = I(\vec{x}) \frac{d^\alpha}{(z(\vec{x})+d)^\alpha}
\end{equation}
where $d$ is the distance of the light source to the Xerox screen. For a point source of light $\alpha=2$, while for a long bar of light $\alpha=1$. By including a factor of $d^\alpha$, we have chosen the convention that the image is not darkened when $z=0$. However, this factor is largely arbitrary and its choice only affects the dynamic range of the darkened image, which can be easily adjusted at later stages.

\section{Recovery (Inverse Problem)} \label{inverse}
We now present a formulation of the ``inverse problem,'' which concerns the recovery of the original image from a projected result (fig. \ref{fig:forward_monet_b}, \ref{fig:forward_monet_c}, \ref{fig:real_scan}). We assume knowledge of the membrane shape, so that our problem amounts to recovering the image as it would appear on a flat book page, given only the scanned version and the shape of the curved page.

\subsection{Undistortion}
To invert the differential distortion, note that the distortion factor from the projected to original image is the multiplicative inverse of that from the original to the projection. Therefore the distortion can be inverted via

\begin{align}
    I(\hat{s},y) &= \textrm{Dist}(I)(x, y), \\
    \frac{d\hat{s}}{dx} &= \frac{1}{\cos \theta (x)}.
\end{align}

\subsection{Deblurring}
To invert the blur, one can apply the inverse of the linear kernel defined in equation \ref{blur_kernel}:

\begin{align}
    I &= \textrm{Blur}^{-1}(B(\vec{x})) \\
    &= \int d^2\vec{x}' \; K^{-1}(\vec{x},\vec{x}') B(\vec{x}') .
\end{align}
In practice it is favorable to avoid calculating $ K^{-1}$ directly, instead making use of numerical methods to solve the relevant linear algebraic equation to reduce computational expense (see section \ref{inverse_discrete} of the supplementary material for details).

\subsection{Lightening}
To invert the darkening, one multiplies the image by a factor proportional to a positive power of the distance from the light source,

\begin{equation}
    \textrm{Dark}^{-1}(I(\vec{x})) \propto \left(z(\vec{x})+d\right)^\alpha I(\vec{x}),
\end{equation}
where the factor of proportionality is now considered arbitrary.

\begin{figure*}[t!]
        \centering
    \begin{subfigure}[t]{0.45\textwidth}
        \centering
        \includegraphics[height=1.8in]{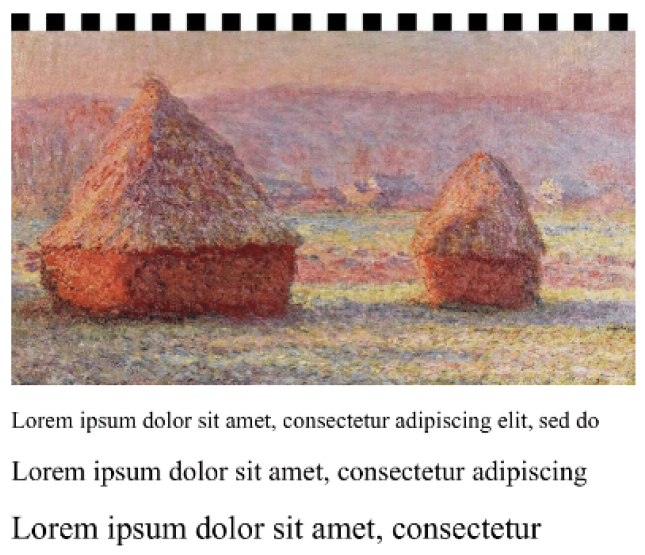}
        \caption{}
        \label{fig:forward_monet_a}
    \end{subfigure}
    \begin{subfigure}[t]{0.45\textwidth}
        \centering
        \includegraphics[height=1.8in]{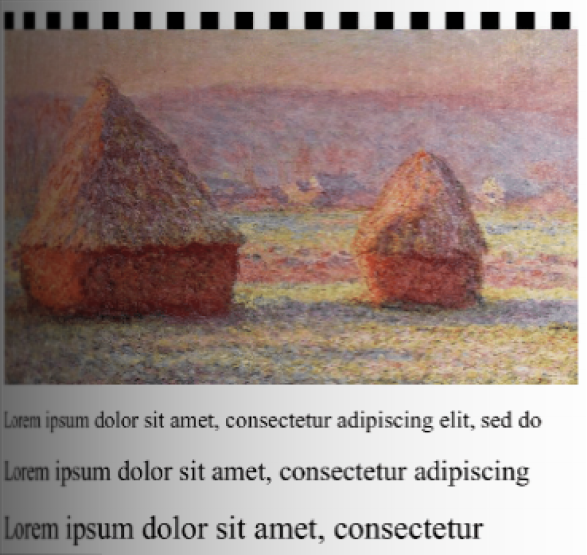}
        \caption{}
        \label{fig:forward_monet_b}
    \end{subfigure}
    \begin{subfigure}[t]{0.45\textwidth}
        \centering
        \includegraphics[height=1.8in]{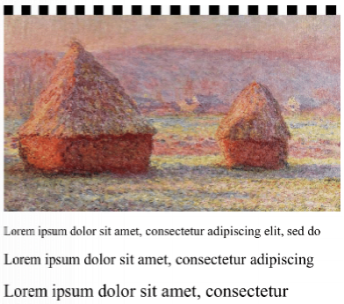}
        \caption{}
        \label{fig:forward_monet_c}
    \end{subfigure}
    \begin{subfigure}[t]{0.45\textwidth}
        \centering
        \includegraphics[height=2.18in]{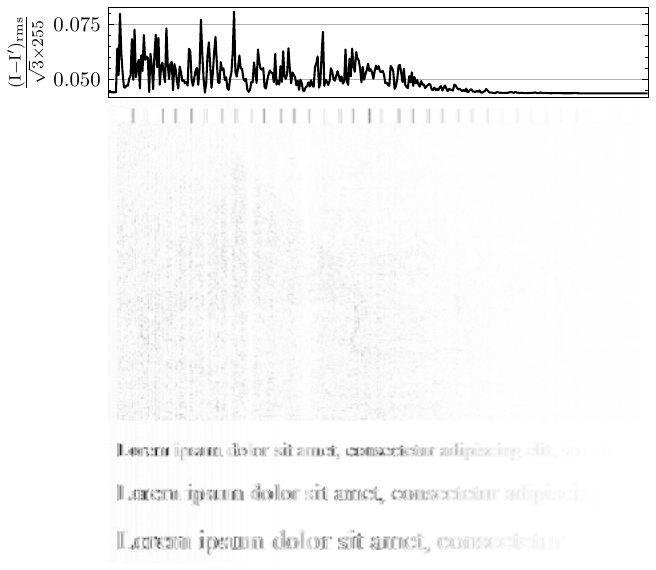}
        \caption{}
    \end{subfigure}
    
    \caption{Implementation of the forward and inverse problems (see section \ref{discrete} for computational details). a) shows the original image, b) shows the image after applying the simulated projection artifacts (section \ref{forward}), and c) after inverting these artifacts (section \ref{inverse}). d) shows the  difference between the recovery $I^\prime$ and the original $I$, with the above plot showing the root-mean-square of the difference. The average is performed over each column and over color values. Reference image: \textit{Grainstacks, White Frost Effect} by Claude Monet \cite{monet_grainstacks_1889}.}
    \label{fig:forward_monet}
\end{figure*}

\begin{figure*}
        \centering
    \begin{subfigure}[t]{0.3\textwidth}
        \centering
        \includegraphics[height=1.3in]{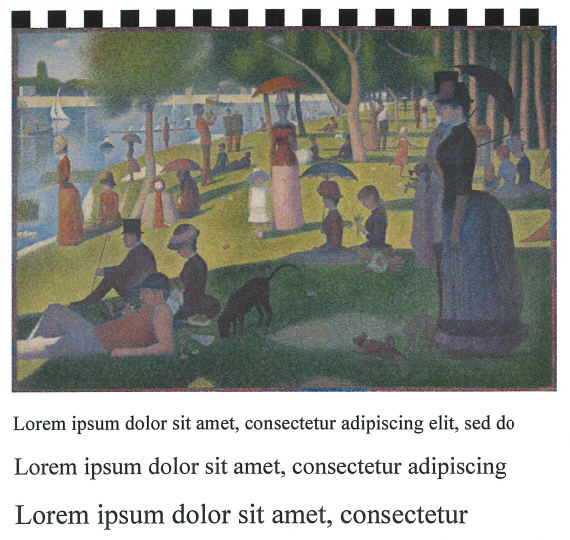}
        \caption{}
        \label{fig:real_scan_a}
    \end{subfigure}
    \begin{subfigure}[t]{0.3\textwidth}
        \centering
        \includegraphics[height=1.3in]{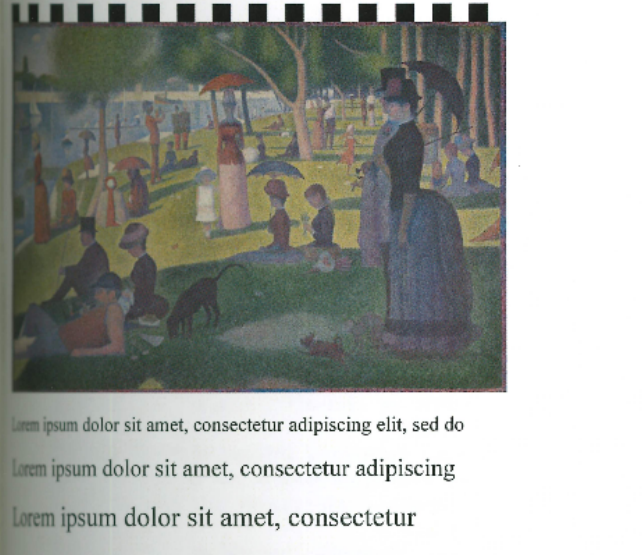}
        \caption{}
        \label{fig:real_scan_b}
    \end{subfigure}
    \begin{subfigure}[t]{0.3\textwidth}
        \centering
        \includegraphics[height=1.3in]{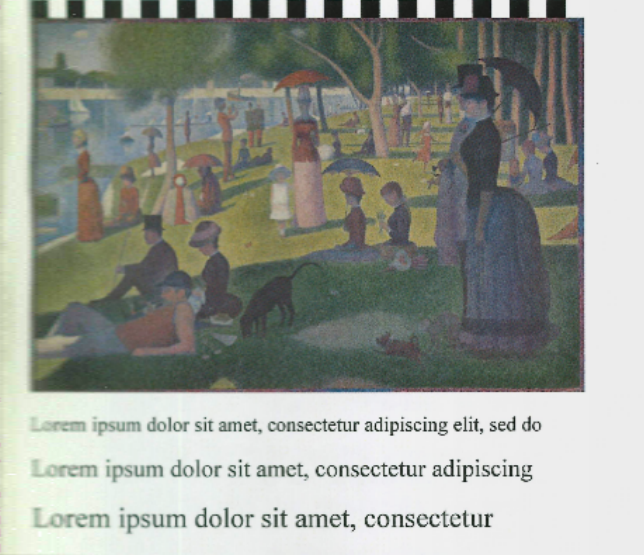}
        \caption{}
        \label{fig:real_scan_c}
    \end{subfigure}
    
    \caption{Application of the recovery algorithm to a real scan. a) shows the original scanned image, while b) shows the recovery generated with parameters $l=66$ pixels, $\theta_0=0.63$ radians. For this application the optimal parameters were determined in a data-driven manner, by measuring the darkening of whitespace and assuming a uniform prior in these areas. Reference image: \textit{A Sunday Afternoon on the Island of La Grande Jatte} by Georges Seurat \cite{seurat_sunday_1884}.} 
    \label{fig:real_scan}
\end{figure*}

\section{Discussion}

Our approach to the humdrum but ubiquitous problem of recovering an image from a xerox scan is predicated on how to parameterize the shape of the scanned page as well as the behavior of the scanner via physical models of few parameters. Our model relies only on physical parameters of the book page and scanner, and can therefore be applied regardless of the nature of the page's content. This leads to a vast reduction in dimensionality of the solution space by imposing constraints corresponding to physical feasibility within the more general paradigm of limiting data-driven models. While our results are promising, ideally one would like to connect our physically motivated model with data-driven machine learning techniques for efficient optimization and exploration within this reduced parameter space, avoiding resource-intensive exploration of a higher-dimensional space comprising mostly unphysical and therefore undesirable models. 

\label{data_driven}

\begin{figure*}
    \centering
    \begin{subfigure}[t]{0.45\textwidth}
        \centering
        \includegraphics[height=1.5in]{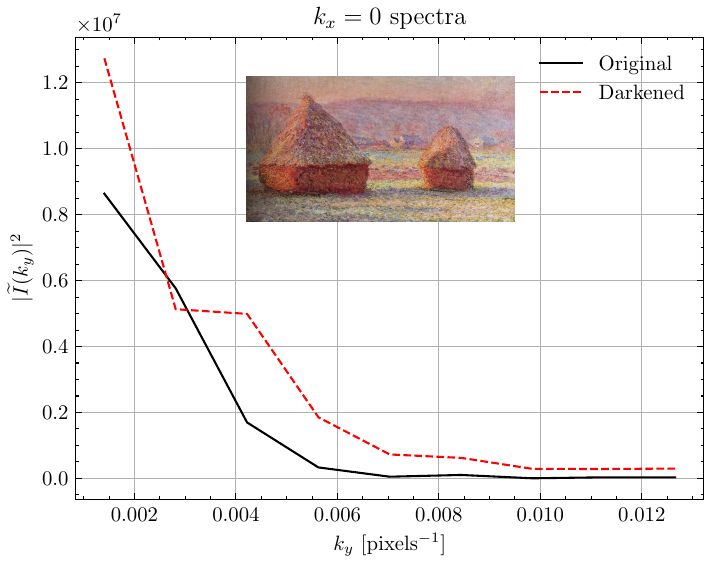}
        \caption{}
        \label{fig:hor_corr_a}
    \end{subfigure}
        \centering
    \begin{subfigure}[t]{0.45\textwidth}
        \centering
        \includegraphics[height=1.5in]{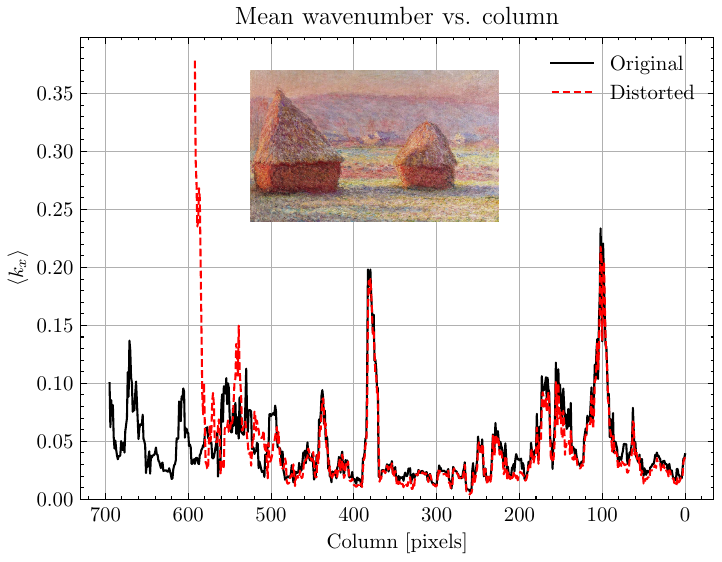}
        \caption{}
        \label{fig:hor_corr_b}
    \end{subfigure}
    \begin{subfigure}[t]{0.45\textwidth}
        \centering
        \includegraphics[height=1.6in]{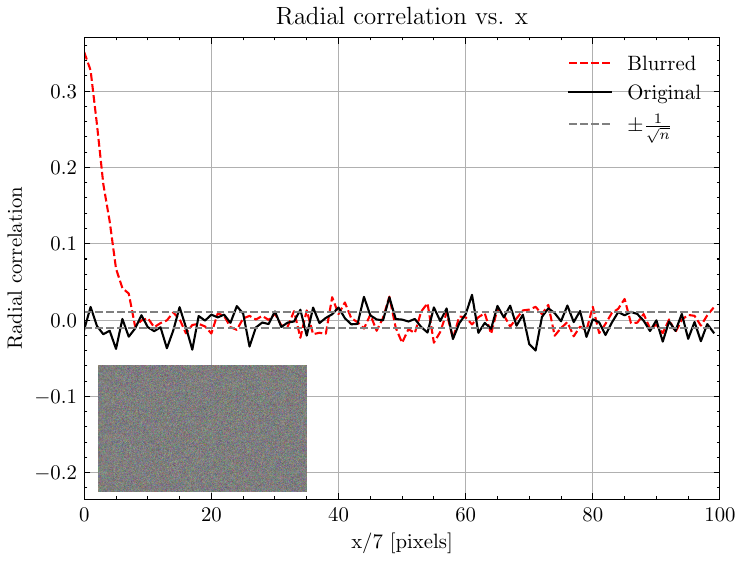}
        \caption{}
        \label{fig:hor_corr_c}
    \end{subfigure}
    \begin{subfigure}[t]{0.45\textwidth}
        \centering
        \includegraphics[height=1.6in]{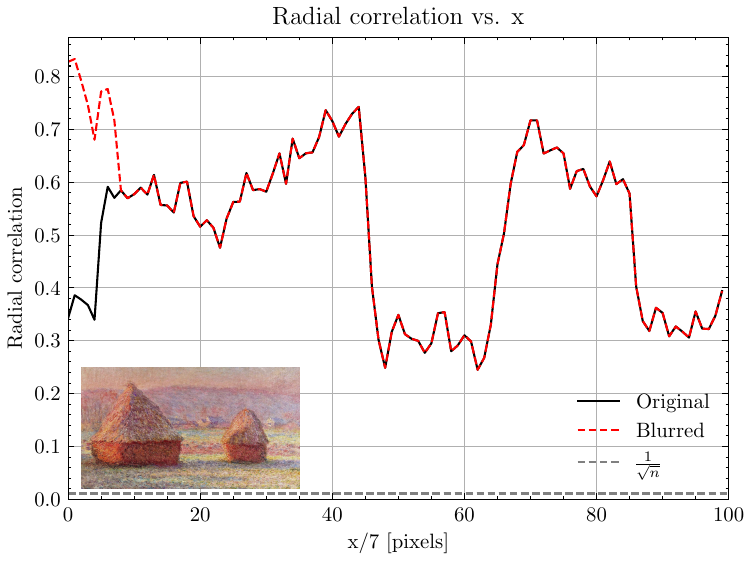}
        \caption{}
        \label{fig:horr_corr_d}
    \end{subfigure}
    
    \caption{Various data-driven methods for detecting artifacts in scanned images. The insets show the image containing each considered artifact. a) shows that darkening increases the absolute values of low-wavenumber Fourier components, b) shows the increase in mean wavenumber due to distortion, and c) and d) show the increase in radial correlations between pixel values due to blurring. In c) the blur is easily detectable, as for a random image one expects correlations of order $\frac{1}{\sqrt{n}}$ without blurring, as much higher correlations with blurring. In d) this method is less successful, since the structure of the image makes it difficult to tell when correlations are caused by blurring versus the features of the image.}
    \label{fig:abstract_prior}
\end{figure*}

In reality one is interested in applying an image recovery algorithm such as this one to real scans of images (fig. \ref{fig:real_scan}) rather than the simulated scans we have discussed. In such applications, one may not know the correct length scale $l$ and boundary angle $\theta_0$ to use for the physical model, nor the length scales $\frac{z}{\sigma}$ and $d$ associated with the optical elements of the scanner. When these parameters are unknown, image recovery can only take place by iterating over a possible range of parameters and comparing the output of the recovery algorithm to a prior which describes one's expectation of the original image, i.e. the ``correct'' recovery. Of course, such algorithms become extraneous in applications where this prior is already precisely known. 

Our physical model of a book page is liable to break down very close to the spine of a book, where the binding may be sufficiently complex to cause curvatures other than those considered. The weight of other pages as well as the book's cover is also not accounted for, and is expected to cause the page's shape to depart from our model where it contacts the scanned page (fig. \ref{fig:shape_verification_2}). 

Furthermore, the optics of the scanner may cause our simple scanning model to be incorrect. For example, the blurring may be nonlinear in the distance from the focal plane, the image may be distorted in the $y$-direction depending on the direction in which the scan is performed, and there may exist nonlinearities in the scanner's photometric response that cause deviations from our model of darkening \cite{mullikin_methods_1994}. We also note that there are various other potential types of errors associated with flatbed scanners that our model does not address \cite{baltsavias_test_1994}. 

We therefore suggest the following approaches to be used in combination with our physical model:

\begin{itemize}
    \item Using abstract or high-level priors that reflect qualitative expectations of the original image (fig. \ref{fig:abstract_prior}), combined with a data-driven approach to find the parameters which best match the recovery with these abstract priors. We have applied such an approach for the recovery depicted in figure \ref{fig:real_scan}; such approaches are difficult on their own due to the complication of translating qualitative expectations to quantitative priors, but can be greatly simplified when combined with our physical model.
    \item Including calibration areas in each scanned page to assist in determining the best parameters, or including calibration areas in a few scanned pages and interpolating the best parameters for pages in between these (fig. \ref{fig:shape_verification_serial}). An example of such calibration areas are included at the tops of the images featured in figures \ref{fig:forward_monet} and \ref{fig:real_scan}.
    \item Applying data-driven machine learning techniques such as convolutional neural networks combined with training data in the form of images lacking any scanning artifacts. In other words, learning a black-box prior against which to compare the results of the recovery algorithm for selection of the correct parameters \cite{schuler_machine_2013, li_blind_2019}. 
\end{itemize}

\section{Acknowledgments}
We thank Yeonsu Jung for assistance with the experiment, and the members of the Mahadevan group for useful discussions. This material is based upon work supported by the National Science Foundation Graduate Research Fellowship under Grant No. DGE 2140743, the Simons Foundation, and the Henri Seydoux Fund.

\bibliographystyle{RS}    
\bibliography{references} 

\section{Supplementary Material}
\setcounter{figure}{0}
\renewcommand{\thefigure}{S\arabic{figure}}

\subsection{Discrete Implementation} \label{discrete}
In this section we outline a discrete implementation of all steps of the forward and inverse problems, as well as provide example Python code which implements each step in a simplified manner. Such an implementation is necessary when applying the algorithm to digitized image scans, which are defined over a discrete domain. We also keep track of the various parameters introduced, which amount to three length scales and an angle, or equivalently 3 dimensionless parameters.

\subsubsection{Physical Model}
We begin by parameterizing the shape of the membrane using our physical model. The only parameters of this step are $l$ and $\theta_0$; from these $\theta(\hat{s})$, $x(\hat{s})$, and $z(\hat{s})$ can be calculated as specified via \ref{theta_s} and \ref{x_s}. In practice one defines a grid to represent $\hat{s}$, then each element of this grid is passed through \ref{theta_s} and \ref{x_s} to define grids describing $\theta(\hat{s})$, $x(\hat{s})$, and $z(\hat{s})$; the horizontal pixel count of the image provides a natural size for these grids (C\ref{implementation_phys}).

We label the grid describing $\hat{s}$ as $s_i$, where $i=1,\dots,N$, and the other grids similarly ($x_i = x(s_i)$, etc.). The original image will thus be a matrix $I_{ij} = I(s_i, y_j)$.

\begin{code}
\begin{mintedbox}{python}
theta_0 = 0.25*np.pi
l = 1
gridsize = input_image.shape[1] # n_columns
s_max = 5
ds = s_max / gridsize

s = np.linspace(0, s_max, gridsize) # s hat
theta = 4.0 * np.arctan(np.exp(-s) 
                        * np.tan(theta_0 / 4.0))
x = l * np.cumsum(np.cos(theta)) * ds
z = l * np.cumsum(np.sin(theta)) * ds
\end{mintedbox}
\caption{Example Python code for parametrization of a page's shape according to physical model. Note that $\hat{s}$ is defined along an evenly spaced grid, while the other variables are not evenly spaced.}
\label{implementation_phys}
\end{code}

\subsubsection{Forward Simulation}
\paragraph{Distortion}\mbox{}\\
To implement local distortion, one can apply \ref{distort} in a discrete manner by using the grids defined for $\hat{s}$, $x$, and $\theta$:
\begin{align}
\label{dist_discrete}
    \textrm{Dist}(I)(x_i, y_j) &= I(s_i, y_j) = I_{ij}, \\
    \frac{x_{i+1}-x_i}{s_{i+1}-s_i}  &= l \cos \theta_i .
\end{align}
The main difficulty one encounters when discretizing this step is the loss of continuity in the domain of $I$; in general the values of $\hat{s}$ derived from \ref{distort} will not be located on the predefined grid for $\hat{s}$. To resolve this, one can either apply an interpolation scheme (C\ref{implementation_dist}) or define another grid with a larger size. The latter method is more computationally expensive but can be made lossless, whereas the first method will induce information loss from the finite grid size. Such loss will be greatest near the spine, where the distortion factors are highest, and introduces imperfection in the image recovery process.

\begin{code}
\begin{mintedbox}{python}
dist_image = np.zeros(input_image.shape)
dx = (x.max() - x.min()) / (gridsize - 1.0)
x_even = np.linspace(x.min(), x.max(), gridsize)
    
# at each x, interpolate desired s values
for i in range(gridsize):
    # nearest-neighbor
    i_near = (np.abs(x - x_even[i])).argmin()
    x_near = x[i_near]
    
    # linear interpolation between two neighbors
    if x_near >= x_even[i]:
        dist_image[:,i] = input_image[:,i_near] 
            * (1.0 - (x_near - x_even[i]) / dx)
            + input_image[:,i_near-1]
            * (x_near - x_even[i]) / dx
    else:
        dist_image[:,i] = input_image[:,i_near+1]
            * (x_near - x_even[i]) / dx
            + input_image[:,i_near]
            * (1.0 - (x_near - x_even[i]) / dx)        
\end{mintedbox}
\caption{Example Python code for applying forward problem of distortion via a lossy but simple natural-neighbor interpolation method.}
\label{implementation_dist}
\end{code}

\paragraph{Blurring}\mbox{}\\
A spatially-dependent blurring can be implemented by applying a discretized version of \ref{blur_kernel}, with the integral replaced by a sum and the kernel $K(\vec{x}, \vec{x}')$ replaced with a tensor, 
\begin{align}
    &\textrm{Blur}(I_{ij}) = \sum_{lm} K_{ijlm} I_{lm}, \\
    K_{ijlm} &= \frac{1}{\sigma(z(x_l)) \sqrt{2\pi}} \exp\left( \frac{-(x_i-x_l)^2-(y_j-y_m)^2}{2 \sigma(z(x_l))^2} \right).
\end{align}
It is also sensible to unravel $I_{ij}$ into a vector $I_\mu$, in which case $K_{ijlm}$ becomes a matrix $K_{\mu \nu}$. Such a redefinition of indices will be useful for formulating deblurring as a matrix inversion. Note that to calculate $K_{ij}$, the function $\sigma(z)$ must be defined, which necessitates the introduction of additional length scales (C\ref{implementation_blur}). Since we assume linear optics with a focal plane at $z=0$, we have $\sigma(z) \propto z$, and so defining $\sigma(z)$ requires one length scale, namely $\frac{z}{\sigma}$.

\begin{code}
\begin{mintedbox}{python}
k = np.zeros((gridsize,)*4)
sigma_l = 1 # blur length scale

for i in range(gridsize):
    for j in range(gridsize):
        sigma = z[j] / sigma_l
        y_mesh, x_mesh 
            = np.mgrid[input_image.shape[0], 
                gridsize]
        center = y[i], x[j]
        dist = np.sqrt((y-y_mesh)**2+(x-x_mesh)**2)
        
        k[i,j,:,:] = gauss(dist, sigma)

blur_image = np.einsum('ijkl,kl->ij', k, input_image)
\end{mintedbox}
\caption{Example Python code for applying forward simulation of blur. \texttt{gauss} is assumed to be a predefined Gaussian function.}
\label{implementation_blur}
\end{code}

\paragraph{Darkening}\mbox{}\\
Implementing darkening requires elementwise multiplication of $I$ with a darkening mask defined by \ref{darken} (C\ref{dark_sim}):
\begin{equation}
\label{dark_discrete}
    \textrm{Dark}(I_{ij}) \propto \frac{I_{ij}}{(z_i + d)^\alpha}.
\end{equation}
Note that this requires the introduction of the length scale $d$ (fig. \ref{fig:book_membrane}).

\begin{code}
\begin{mintedbox}{python}
d = 1 # darkening length scale

dark_mask = 1.0 / (z + d)**2

dark_image = dark_mask * input_image
\end{mintedbox}
\caption{Python code for applying forward simulation of darkening.}
\label{dark_sim}
\end{code}

\subsubsection{Inverse Recovery} \label{inverse_discrete}
\paragraph{Undistortion}\mbox{}\\
To recover the undistorted image, we can apply the inverse of \ref{dist_discrete}:

\begin{align}
     I(s_i, y_j) &= \textrm{Dist}(I)(x_i, y_j), \\
    \frac{s_{i+1}-s_i}{x_{i+1}-x_i}  &= \frac{1}{l \cos \theta_i}.
\end{align}
Note that this method will perform poorly where the distortion factor $\cos\theta$ is high due to the finite resolution of the image scan. Specifically, if several features are projected onto only one pixel, these features will be unrecoverable. 

\paragraph{Deblurring} \label{deblurring_discrete}
\mbox{}\\
To invert blur, one must invert the blur kernel $K_{ijlm}$. For this case it is helpful to redefine $I$ as a vector $I_{\mu}$ and $K$ as a matrix $K_{\mu \nu}$ as previously discussed. Then the deblurring formally proceeds as (C\ref{deblur_sim})

\begin{equation}
    I_{\mu} = \sum_\nu \left( K^{-1} \right)_{\mu \nu} \textrm{Blur}(I)_{\nu} .
\end{equation}

It should be noted that inverting $K$ in this manner will be expensive and difficult due to the large size and ill-conditioning of the kernel $K$. Therefore it is advisable to instead implement deblurring by solving the linear algebraic equation

\begin{equation}
    \sum_\nu K_{\mu \nu} I_{\nu} =  \textrm{Blur}(I)_{\mu} .
    \label{matrix_eq}
\end{equation}
for $I_\nu$, where $\textrm{Blur}(I)_{\mu}$ and $K_{\mu \nu}$ are known. For such a problem various methods can be used that address the ill-conditioned nature of $K$, such as Tikhonov regularization or truncated singular value decomposition. Furthermore, many of the values of $K$ will be extremely small, and so one can introduce a cutoff, e.g. for elements of $K$ corresponding to $|\vec{x}-\vec{x}'|>5\sigma(z(\vec{x}'))$, below which such small elements are set to zero. Such an approximation introduces minimal error in the blurring kernel and enables one to use sparse matrix methods such as sparse Gaussian elimination or sparse LU decomposition in the numerical solution of equation \ref{matrix_eq} \cite{press_numerical_2007}.

\begin{code}
\begin{mintedbox}{python}
k = k.reshape(n, n)

k_inv = np.linalg.pinverse(k) # Likely to fail

deblur_image = np.einsum('uv,v->u', k, blur_image)
\end{mintedbox}
\caption{Example Python code for applying deblurring, assuming $K$ is defined as in Code \ref{implementation_blur}. Note that \mintinline{latex}{np.linalg.pinverse} is likely a poor inversion method due to the ill-conditioning of $K$; in a real application one should employ numerical techniques for solving linear algebraic equations involving sparse, ill-conditioned matrices (section \ref{discrete}\ref{deblurring_discrete}).}
\label{deblur_sim}
\end{code}

\paragraph{Lightening}\mbox{}\\
To lighten the image one can apply the inverse of the darkening mask in equation \ref{dark_discrete} (C\ref{light_sim}):

\begin{equation}
    I_{ij} \propto \textrm{Dark}(I_{ij}) \cdot (z_i + d)^\alpha .
\end{equation}

\begin{code}
\begin{mintedbox}{python}
d = 1 # darkening length scale

light_mask = 1.0 * (z + d)**2

light_image = light_mask * proj_image
\end{mintedbox}
\caption{Example Python code for applying inverse recovery of darkening.}
\label{light_sim}
\end{code}

\begin{figure}[hbt!]
    \centering
    \includegraphics[height=3.0in]{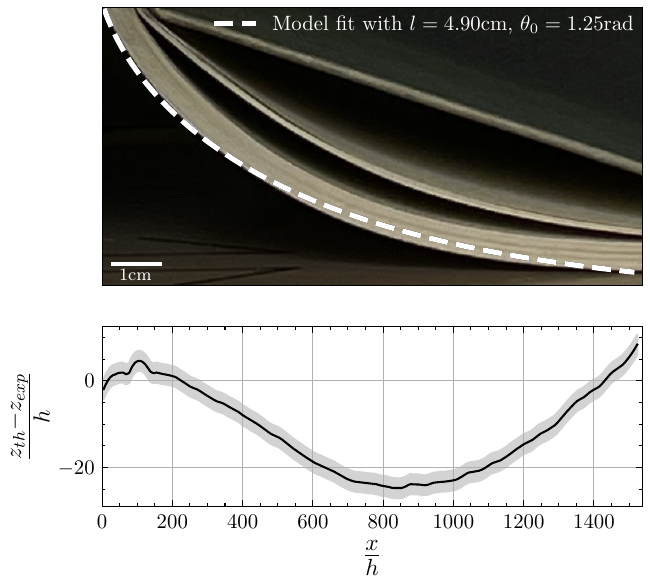}
    \caption{A case where the physical model performs poorly due to the contact between of the scanned page with neighboring pages and the book cover. However, note that the restoration which assumes such a page shape will still yield (incomplete) artifact reduction.} 
    \label{fig:shape_verification_2}
\end{figure}

\begin{figure}
    \centering
    \includegraphics[height=3.0in]{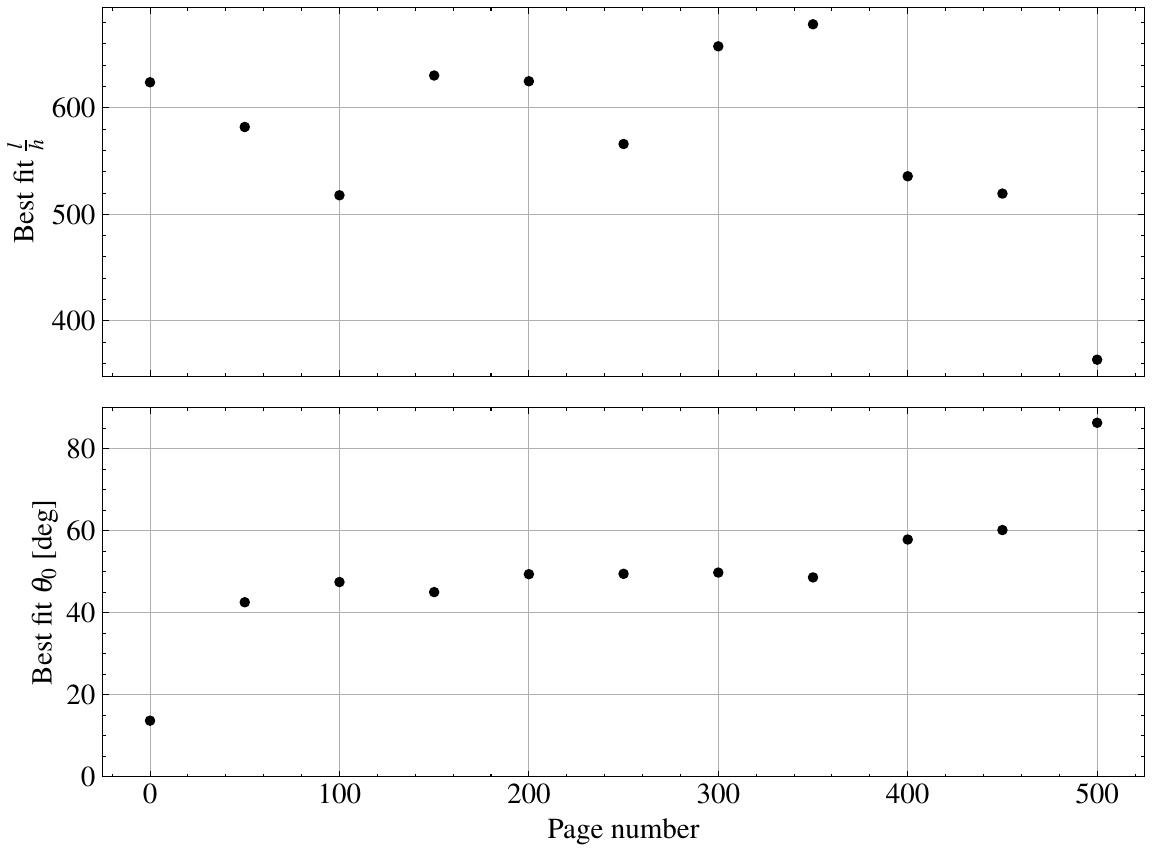}
    \caption{Fitting the physical model to various pages from the same book as in fig. \ref{fig:shape_verification}. We see a systematic change in $\theta_0$, but no such change in $l$.}
    \label{fig:shape_verification_serial}
\end{figure}

\end{document}